\documentclass[reprint,showpacs,preprintnumbers,nofootinbib,amsmath,amssymb,aps,prd,floatfix]{revtex4-1}
\pdfoutput=1
\usepackage{graphicx}
\usepackage{bm}
\usepackage{subfigure}
\usepackage{url}
\usepackage[hyperindex]{hyperref}
\usepackage{color}
\usepackage[ddmmyy,24hr]{datetime}
\usepackage{bigdelim}
\usepackage{booktabs}
\usepackage{dcolumn}
\usepackage{multirow}
\usepackage{subfigure}
\usepackage{cancel}
\usepackage{stackrel}
\usepackage{paralist}
\usepackage{xspace}
\newcommand{\nua}[1]{\ensuremath{\rlap{\kern-2.5pt\ensuremath{\overset{\scriptscriptstyle(-)}{\phantom{\nu}}}}{\ensuremath{{\nu}_{#1}}}}}

\usepackage{enumitem}

\makeatletter
\def\namedlabel#1#2{\begingroup
    #2%
    \def\@currentlabel{#2}%
    \phantomsection\label{#1}\endgroup
}
\makeatother
\begin{document}

\title{Model-Independent $\bar\nu_{e}$ Short-Baseline Oscillations from Reactor Spectral Ratios}

\author{S. Gariazzo}
\affiliation{Instituto de F\'isica Corpuscular (CSIC-Universitat de Val\`encia), Paterna (Valencia), Spain}

\author{C. Giunti}
\affiliation{Istituto Nazionale di Fisica Nucleare (INFN), Sezione di Torino, Via P. Giuria 1, I--10125 Torino, Italy}

\author{M. Laveder}
\affiliation{Dipartimento di Fisica e Astronomia ``Galileo Galilei'', Universit\`a di Padova, Via F. Marzolo 8, I--35131 Padova, Italy}
\affiliation{Istituto Nazionale di Fisica Nucleare (INFN), Sezione di Padova, Via F. Marzolo 8, I--35131 Padova, Italy}

\author{Y.F. Li}
\affiliation{Institute of High Energy Physics,
Chinese Academy of Sciences, Beijing 100049, China}
\affiliation{School of Physical Sciences, University of Chinese Academy of Sciences, Beijing 100049, China}

\date{16 April 2018}

\begin{abstract}
We consider the ratio of the spectra measured in the DANSS neutrino experiment
at 12.7 and 10.7~m from a nuclear reactor.
These data give a new model-independent indication
in favor of short-baseline $\bar\nu_{e}$ oscillations
which reinforce the model-independent indication
found in the late 2016 in the NEOS experiment.
The combined analysis of the NEOS and DANSS spectral ratios
in the framework of 3+1 active-sterile neutrino mixing
favor short-baseline $\bar\nu_{e}$ oscillations
with a statistical significance of
$3.7\sigma$.
The two mixing parameters
$\sin^{2}2\vartheta_{ee}$ and $\Delta{m}^{2}_{41}$
are constrained at $2\sigma$ in a narrow-$\Delta{m}^{2}_{41}$ island at
$\Delta{m}^2_{41} \simeq 1.3 \, \text{eV}^2$,
with
$
\sin^{2}2\vartheta_{ee}
=
0.049
\pm 0.023
$ ($2\sigma$).
We discuss the implications of the model-independent NEOS+DANSS analysis
for the reactor and Gallium anomalies.
The NEOS+DANSS model-independent determination of short-baseline $\bar\nu_{e}$ oscillations
allows us to analyze the reactor rates
without assumptions on the values of the main reactor antineutrino fluxes
and the data of the Gallium source experiments with free detector efficiencies.
The corrections to the reactor neutrino fluxes and the Gallium detector efficiencies
are obtained from the fit of the data.
In particular,
we confirm the indication in favor of the need for a recalculation of
the $^{235}\text{U}$ reactor antineutrino flux
found in previous studies assuming the absence of neutrino oscillations.
\end{abstract}


\maketitle

\section{Introduction}
\label{sec:introduction}

Neutrino oscillations revealed the existence of neutrino masses
and are one of the most powerful tools in the search of new physics beyond the Standard Model.
An interesting indication of new physics is given by the
reactor \cite{Mention:2011rk}
and
Gallium \cite{Abdurashitov:2005tb,Laveder:2007zz,Giunti:2006bj,Acero:2007su,Giunti:2009zz,Giunti:2010zu,Giunti:2012tn}
short-baseline neutrino oscillation anomalies,
which can be explained by the existence of a non-standard sterile neutrino
at the eV mass scale
(see the recent fits in
Refs.~\cite{Gariazzo:2017fdh,Dentler:2017tkw}).

The reactor antineutrino anomaly \cite{Mention:2011rk}
was discovered in 2011 as a consequence of a new calculation of the reactor
$\bar\nu_{e}$ fluxes
\cite{Mueller:2011nm,Huber:2011wv}
due to the fissions of
$^{235}\text{U}$,
$^{238}\text{U}$,
$^{239}\text{Pu}$, and
$^{241}\text{Pu}$.
The predicted total rates of $\bar\nu_{e}$ detection
were found to be a few percent larger
than those obtained in previous calculations~\cite{Vogel:1980bk,Schreckenbach:1985ep,Hahn:1989zr}.
The comparison of the new predicted rates with the rates measured in several experiments at
distances between a few meters and about 500 meters from a reactor
indicate a deficit of about 5\%
which can be explained by the disappearance of $\bar\nu_{e}$ during their propagation from the reactor to the detector,
that is most likely due to active-sterile neutrino oscillations
(see the review in Ref.~\cite{Gariazzo:2015rra}).
However,
the correctness of the $\bar\nu_{e}$ flux calculation has been put into question
(see Refs.~\cite{Huber:2016fkt,Hayes:2016qnu})
by the discovery of the so-called ``5 MeV bump''
of the reactor antineutrino spectrum measured
in the
RENO \cite{Seo:2014xei,RENO:2015ksa},
Double Chooz \cite{Abe:2014bwa},
Daya Bay \cite{An:2016srz},
and
NEOS \cite{Ko:2016owz}
experiments.
Moreover,
the Daya Bay measurement \cite{An:2017osx} of the correlation between the
reactor fuel evolution and the antineutrino detection rate indicates that at least
the calculation of the $^{235}\text{U}$ reactor antineutrino flux must be revised
\cite{An:2017osx,Giunti:2017nww,Giunti:2017yid,Gebre:2017vmm}
(see also the recent review in Ref.~\cite{Qian:2018wid}).

The Gallium neutrino anomaly
\cite{Abdurashitov:2005tb,Laveder:2007zz,Giunti:2006bj,Acero:2007su,Giunti:2009zz,Giunti:2010zu,Giunti:2012tn}
is a deficit of $\nu_{e}$ events
measured in the
Gallium radioactive source experiments
GALLEX
\cite{Anselmann:1994ar,Hampel:1997fc,Kaether:2010ag}
and
SAGE
\cite{Abdurashitov:1996dp,Abdurashitov:1998ne,Abdurashitov:2005tb,Abdurashitov:2009tn}.
As explained in Ref.~\cite{Giunti:2012tn},
in the calculation of the Gallium anomaly
the uncertainties of the neutrino-nucleus cross section are taken into account
using the
${}^{71}\text{Ga}({}^{3}\text{He},{}^{3}\text{H}){}^{71}\text{Ge}$
measurement in Ref.~\cite{Frekers:2011zz}.
However,
the efficiencies of the
GALLEX and SAGE
detectors are not known and could have been overestimated.

In this paper we consider the new results
of the DANSS reactor neutrino experiment
presented in Ref.~\cite{DANSS-171201}~\footnote{
The same results have been published in arXiv:1804.04046
after completion of this work.
}.
We will show that the ratio of the spectra measured in the DANSS experiment
at 12.7 and 10.7~m from a nuclear reactor
provide a model-independent indication of
short-baseline $\nua{e}$ oscillations
which
reinforces the model-independent indication
found in the late 2016 in the NEOS experiment~\cite{Ko:2016owz}.

We will show that
the combined analysis of the NEOS and DANSS spectral ratios
allow us to determine the neutrino mixing parameters
in a model-independent way.
In particular, the determination of neutrino oscillations
is independent from the reactor anomaly,
which depends on the comparison of the measured and calculated reactor rates
\cite{Mueller:2011nm,Huber:2011wv},
and from the Gallium anomaly,
which depends on the estimated efficiencies of the GALLEX and SAGE detectors.
This is a remarkable result that raises to a new level the significance of
the indications in favor of short-baseline
$\nu_{e}$ and $\bar\nu_{e}$ disappearance.

Moreover,
the NEOS+DANSS model-independent determination of the neutrino oscillation parameters
allow us to derive the values of the
${}^{235}\text{U}$, ${}^{238}\text{U}$, and ${}^{239}\text{Pu}$
reactor antineutrino fluxes
which are needed to fit the reactor rates and the Daya Bay evolution data
\cite{An:2017osx}
and the efficiencies of the GALLEX and SAGE detectors
that are needed to explain the Gallium anomaly.

We work in the 3+1 framework explained in Ref.~\cite{Gariazzo:2015rra},
which is a perturbation of the standard three-neutrino mixing framework
which explains the oscillations observed in
solar, atmospheric and long-baseline neutrino experiments
(see Refs.~\cite{Esteban:2016qun,Capozzi:2017ipn,deSalas:2017kay}).
We use the notation of Ref.~\cite{Gariazzo:2017fdh},
of which we only remind that short-baseline oscillations depend on the
squared-mass difference
$\Delta{m}^2_{41}$
and
the amplitude of $\nu_{e}$ and $\bar\nu_{e}$ disappearance
can be parameterized by the effective mixing angle $\vartheta_{ee}$ given by
\begin{equation}
\sin^2 2\vartheta_{ee}
=
4
|U_{e4}|^2
\left( 1 -  |U_{e4}|^2 \right)
,
\label{see}
\end{equation}
$U$ is the $4\times4$ unitary mixing matrix
\cite{Bilenky:1996rw}.
Sometimes $\vartheta_{ee}$ is called $\vartheta_{14}$
(see, for example, Ref.~\cite{Dentler:2017tkw}).

In this paper we consider only short-baseline $\nu_{e}$ and $\bar\nu_{e}$ disappearance
experiments.
In particular we do not consider the LSND anomaly
\cite{Athanassopoulos:1995iw,Aguilar:2001ty},
which is a signal of short-baseline $\bar\nu_{\mu}\to\bar\nu_{e}$ appearance
which can be explained in the framework of 3+1 active sterile neutrino mixing
(see Ref.~\cite{Gariazzo:2015rra}).
However,
the strong limits on $\nua{\mu}$ disappearance obtained recently
in the MINOS and MINOS+ experiments~\cite{Adamson:2017uda}
increase to an unacceptable level
the appearance-disappearance tension discussed in many papers
\cite{Okada:1996kw,Bilenky:1996rw,Kopp:2011qd,Giunti:2011gz,Giunti:2011hn,Giunti:2011cp,Conrad:2012qt,Archidiacono:2012ri,Archidiacono:2013xxa,Kopp:2013vaa,Giunti:2013aea,Gariazzo:2015rra,Gariazzo:2017fdh,Dentler:2017tkw}.
Indeed,
adding the MINOS and MINOS+ data to the data considered in our PrGlo17 fit
\cite{Gariazzo:2017fdh},
we obtain an appearance-disappearance parameter goodness-of-fit of
about 0.4\%~\footnote{
In this analysis we used the public MINOS and MINOS+ code~\cite{Adamson:2017uda}
which relies on the neutrino flux prediction of the MINERvA collaboration~\cite{Aliaga:2016oaz}.
With a ``shape analysis'' of the MINOS and MINOS+ data
allowing different free normalizations for the predictions of the charged-current and neutral-current events,
we obtained an appearance-disappearance parameter goodness-of-fit of
about 0.5\%,
which is still too small.
The details of these analyses will be presented elsewhere.}.
This result disfavors the LSND anomaly,
but a definitive conclusion on the LSND $\bar\nu_{\mu}\to\bar\nu_{e}$ signal
will be possible only after its direct test in the
SBN \cite{Antonello:2015lea}
and
JSNS$^2$ \cite{Ajimura:2017fld}
experiments.

The plan of the paper is as follows.
In Section~\ref{sec:DANSS}
we present our analysis of DANSS data and the results of the
NEOS+DANSS combined fit.
In Section~\ref{sec:anomalies}
we compare the results of the
NEOS+DANSS fit with the reactor and Gallium anomalies.
In Section~\ref{sec:nuedis}
we present the results of a model-independent fit of short-baseline
$\nu_{e}$ and $\bar\nu_{e}$ disappearance data.
Finally,
we draw our conclusions
in Section~\ref{sec:conclusions}.

\begin{figure}[!t]
\centering
\includegraphics*[width=\linewidth]{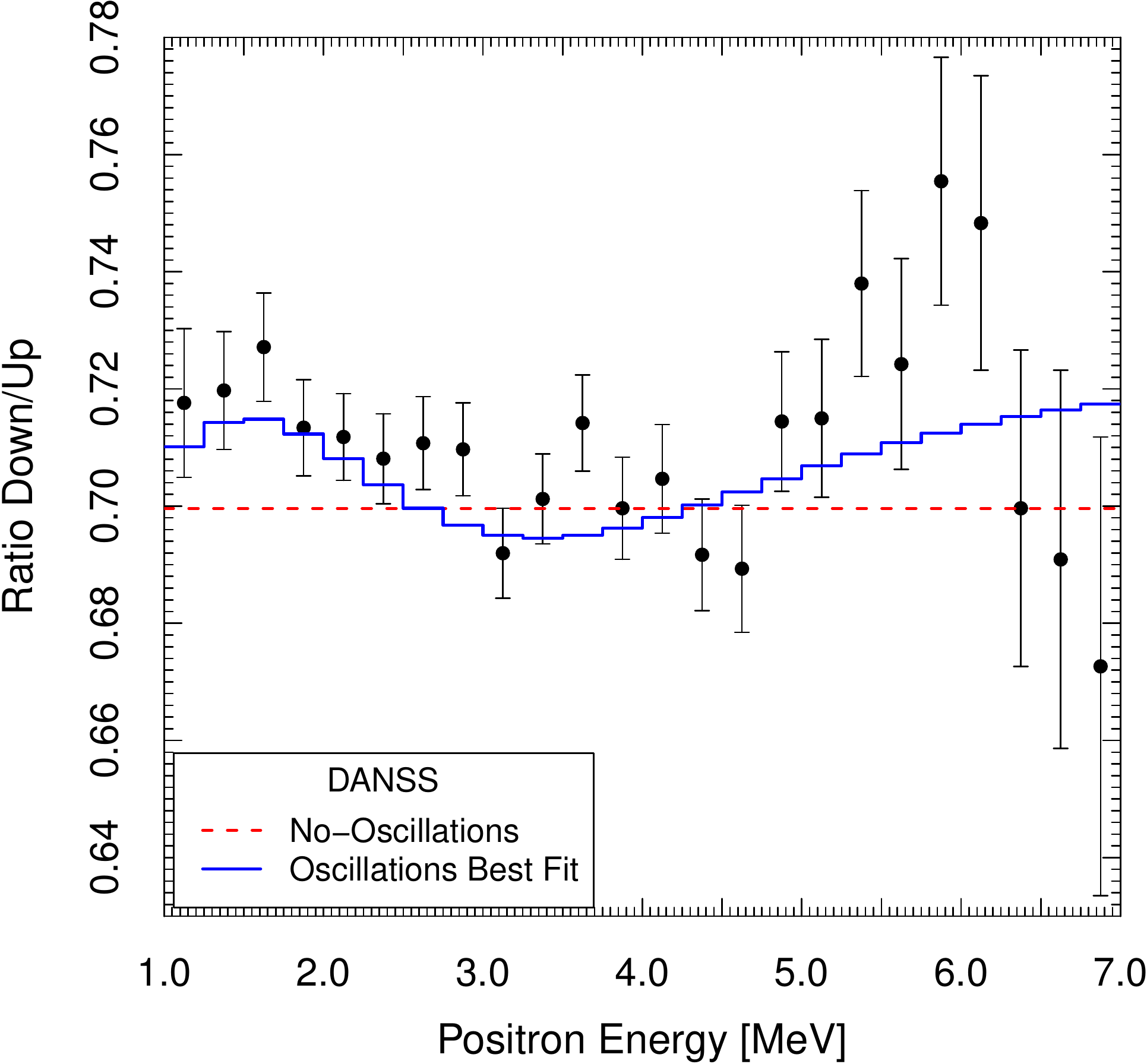}
\caption{ \label{fig:hst-danss}
The Down/Up DANSS spectral data presented in Ref.~\cite{DANSS-171201}.
The red dashed line shows the best-fit of the data without oscillations and a free normalization.
The blues solid line shows the best-fit that we obtained with neutrino oscillations
and a free normalization.
}
\end{figure}

\begin{figure*}[!t]
\centering
\setlength{\tabcolsep}{0pt}
\begin{tabular}{cc}
\subfigure[]{\label{fig:danss}
\includegraphics*[width=0.49\linewidth]{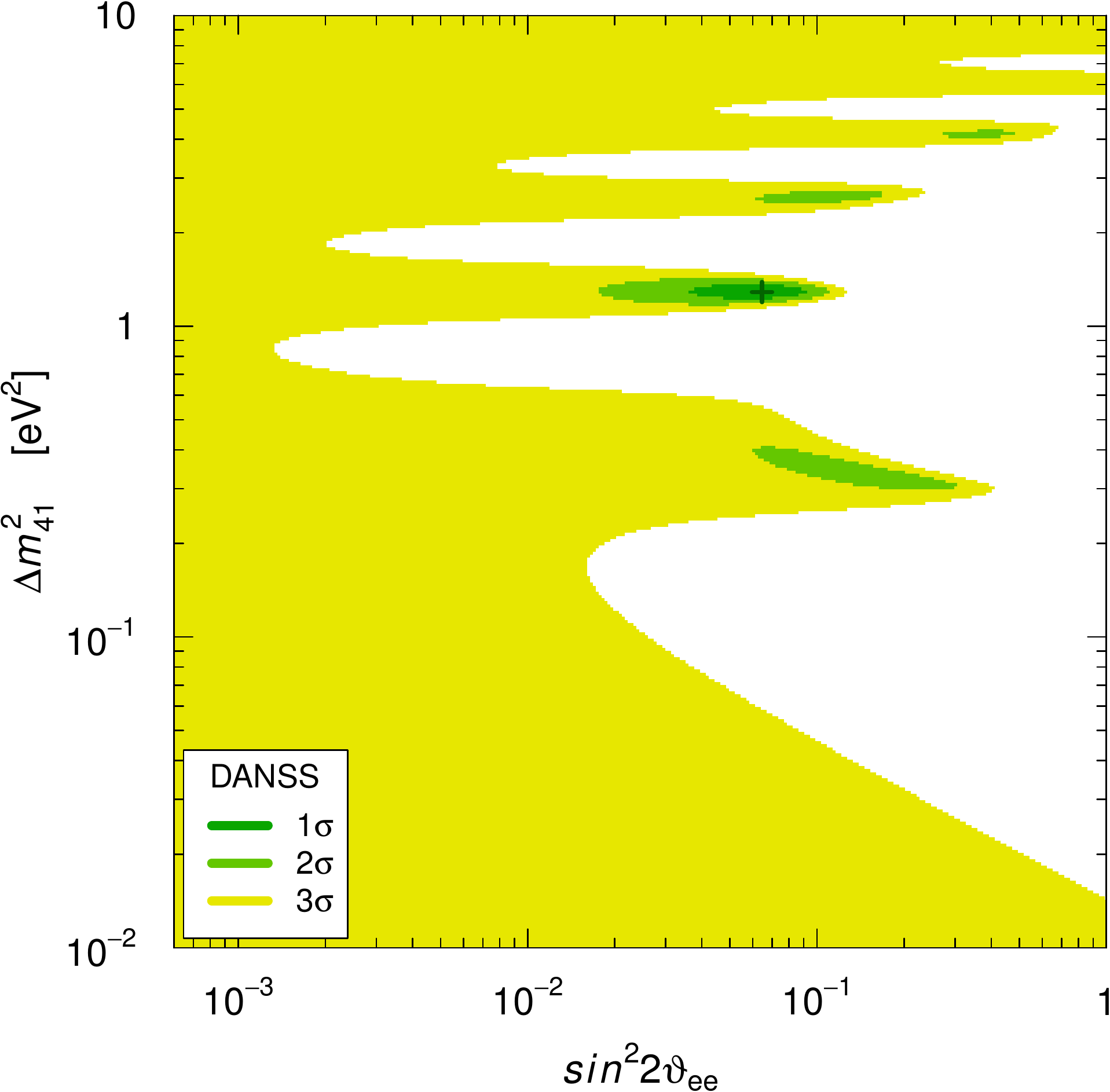}
}
&
\subfigure[]{\label{fig:neos}
\includegraphics*[width=0.49\linewidth]{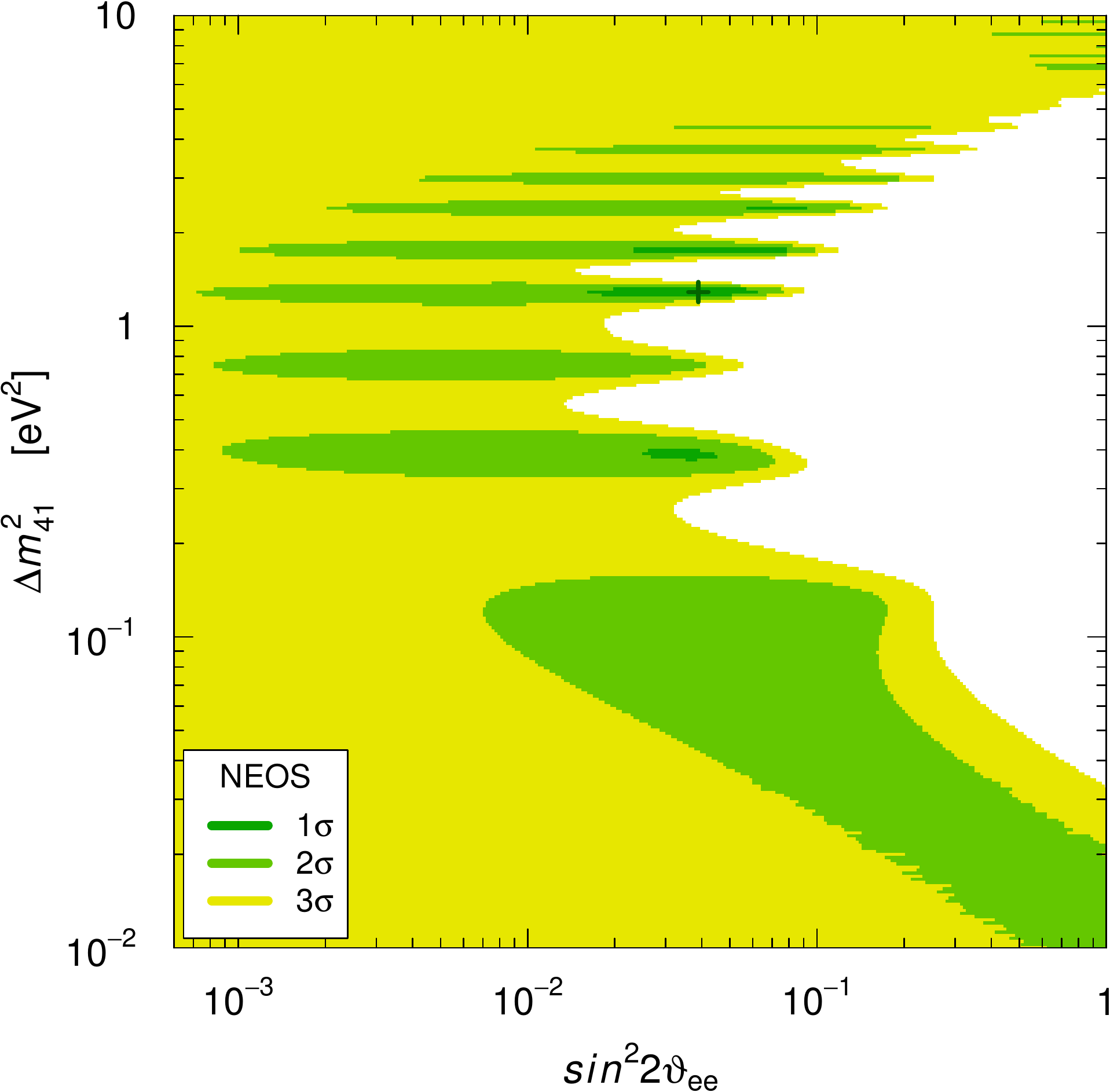}
}
\end{tabular}
\caption{ \label{fig:danss-neos}
Allowed regions in the
$\sin^{2}2\vartheta_{ee}$--$\Delta{m}^{2}_{41}$ plane
obtained from the fits of
\subref{fig:danss}
DANSS~\cite{DANSS-171201}
and
\subref{fig:neos}
NEOS~\cite{Ko:2016owz}
data.
The best-fit points corresponding to the $\chi^2_{\text{min}}$
in Table~\ref{tab:nuedis} are indicated by crosses.
}
\end{figure*}

\begin{figure*}[!t]
\centering
\setlength{\tabcolsep}{0pt}
\begin{tabular}{cc}
\subfigure[]{\label{fig:neos+danss-spe}
\includegraphics*[width=0.49\linewidth]{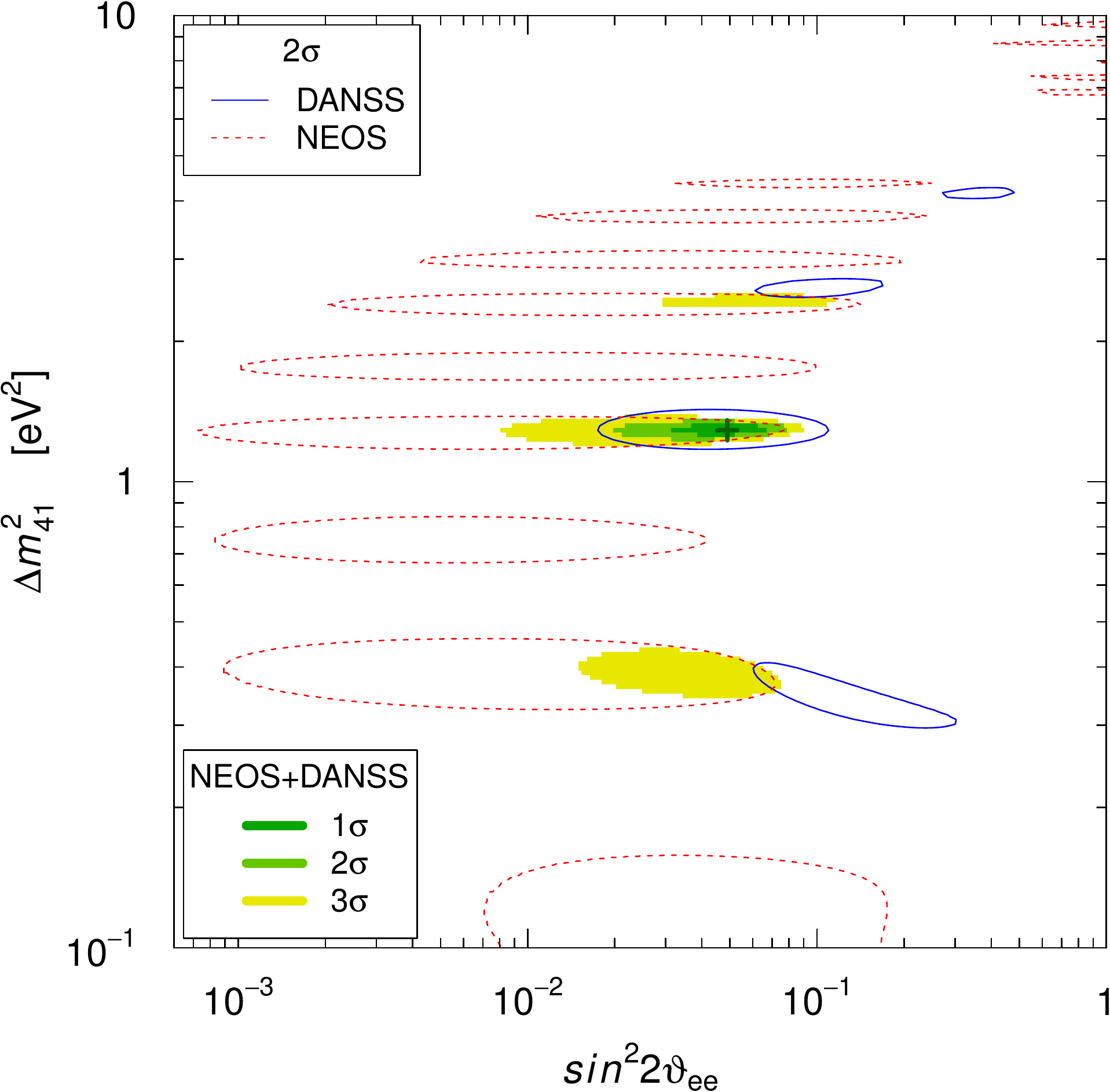}
}
&
\subfigure[]{\label{fig:neos+danss-cmp}
\includegraphics*[width=0.49\linewidth]{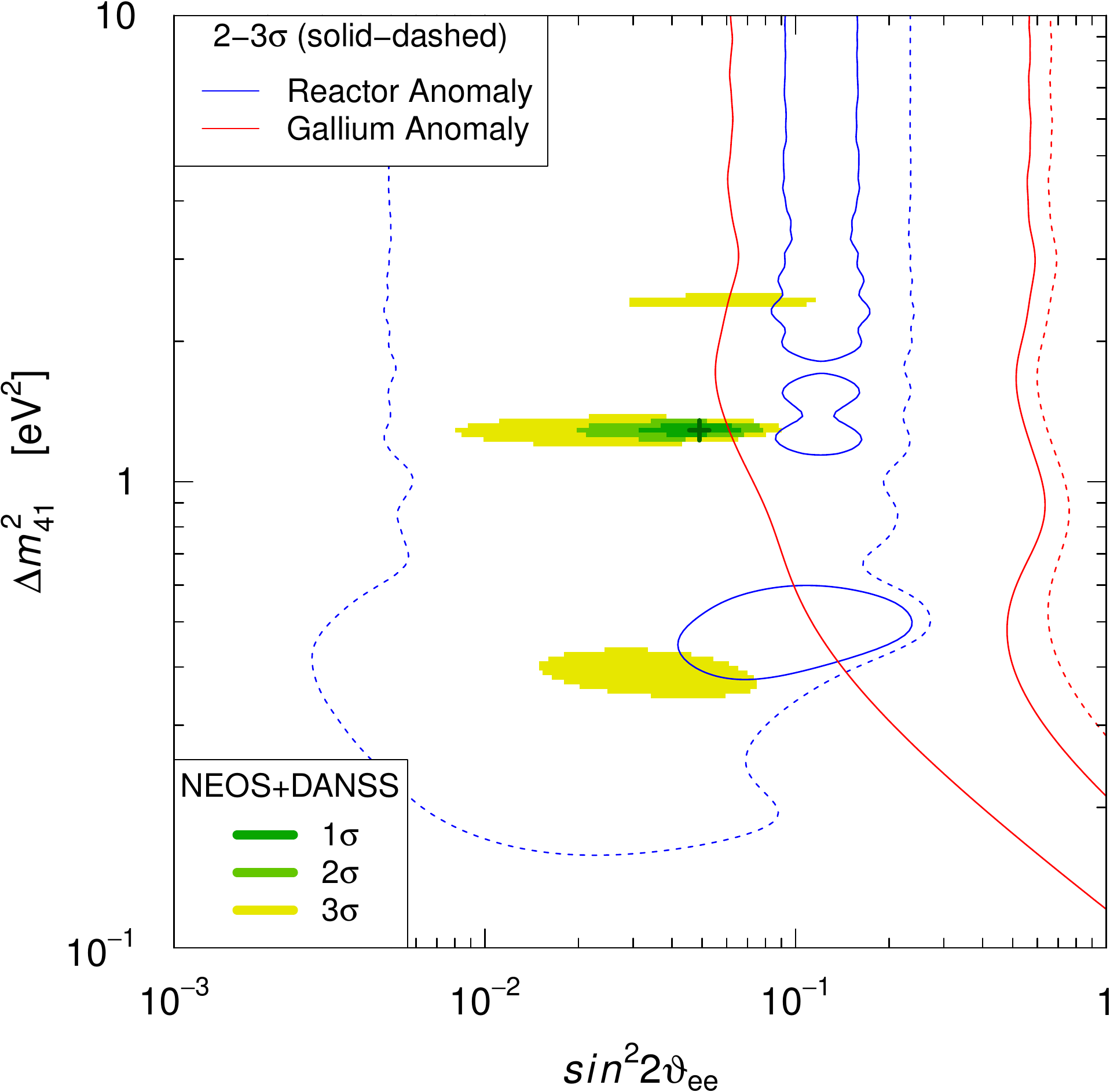}
}
\end{tabular}
\caption{ \label{fig:neos+danss}
Allowed regions in the
$\sin^{2}2\vartheta_{ee}$--$\Delta{m}^{2}_{41}$ plane obtained from
the combined fit of the data of the
DANSS~\cite{DANSS-171201} and NEOS~\cite{Ko:2016owz}
experiments (shaded regions).
\subref{fig:neos+danss-spe}
Comparison of the allowed regions
with the $2\sigma$ allowed regions
of DANSS and NEOS.
\subref{fig:neos+danss-cmp}
Comparison of the allowed regions
with the regions allowed at 2 and 3$\sigma$
by the reactor anomaly and by the Gallium anomaly.
}
\end{figure*}

\section{DANSS and NEOS}
\label{sec:DANSS}

DANSS is a neutrino experiment with a solid scintillator detector
located under a commercial power reactor
which emits a huge $\bar\nu_{e}$ flux leading to a high-statistics measurement.
The DANSS detector is installed on a movable platform which allows to change the
distance between the centers of the reactor and detector from 10.7 to 12.7 m.
We analyzed the Down/Up DANSS data presented in Ref.~\cite{DANSS-171201}
on the ratio of the energy spectra measured at the two distances.

As reported in Ref.~\cite{DANSS-171201},
the DANSS collaboration found that
the best fit of the Down/Up spectral ratio is obtained for short-baseline neutrino oscillations
with a $\chi^2$ that is smaller by 13.3 with respect to the case of no oscillations.
They found the best fit at
$\sin^2 2\vartheta_{ee} = 0.045$
and
$\Delta{m}^2_{41} = 1.4 \, \text{eV}^2$.

The DANSS data are shown in Fig.~\ref{fig:hst-danss}.
In our analysis,
for each energy bin we averaged the oscillation probability over the
geometrical volumes of the reactor and the detector.
We allowed a free normalization of the data
and
we took into account the 25\% energy resolution reported in Ref.~\cite{DANSS-171201}
and a correlated 2\% systematic uncertainty (following Ref.~\cite{Dentler:2017tkw}).
We obtained a $\chi^2$ that is smaller by
11.4
with respect to the case of no oscillations
and the best fit at
$\sin^2 2\vartheta_{ee} = 0.065$
and
$\Delta{m}^2_{41} = 1.3 \, \text{eV}^2$.
These results are in an acceptable approximate agreement with those obtained
by the DANSS collaboration in Ref.~\cite{DANSS-171201} and slightly more conservative.

Figure~\ref{fig:danss} shows the
allowed regions in the
$\sin^{2}2\vartheta_{ee}$--$\Delta{m}^{2}_{41}$ plane
obtained from our analysis of the data of the
DANSS experiment.
It is interesting to compare them with the allowed regions obtained
from the analysis of the
NEOS experiment \cite{Ko:2016owz}
shown in Fig.~\ref{fig:neos}.
One can see that there is a remarkable agreement
between the DANSS and NEOS best-fit regions
which lie in a narrow interval around
$\Delta{m}^2_{41} \simeq 1.3 \, \text{eV}^2$
with compatible values of
$\sin^{2}2\vartheta_{ee}$.

Let us emphasize that these indications in favor of short-baseline neutrino oscillations
are model-independent,
because they depend only on measured spectral ratios:
the Down/Up spectral ratio for DANSS~\cite{DANSS-171201}
and the
NEOS/Daya Bay spectral ratio for NEOS~\cite{Ko:2016owz}
(the NEOS spectrum measured at a distance of 24 m
was normalized to the
Daya Bay spectrum
\cite{An:2016srz}
measured at the large distance of about 550 m,
where short-baseline oscillations are averaged out).
In particular,
these indications do not depend on the calculation of the
reactor $\bar\nu_{e}$ fluxes on which the reactor antineutrino anomaly is based
\cite{Mention:2011rk}.

\begin{table}[!t]
\centering
\renewcommand{\arraystretch}{1.3}
\begin{tabular}[t]{c|}
\\
\\
\hline
$\chi^{2}_{\text{min}}$\\
NDF\\
GoF\\
\hline
$\Delta{m}^2_{41}$\\
$\sin^22\vartheta_{ee}$\\
$r_{235}$\\
$r_{238}$\\
$r_{239}$\\
$\eta_{\text{G}}$\\
$\eta_{\text{S}}$\\
\hline
$|U_{e4}|^2$\\
$\sigma_{f,235}$\\
$\sigma_{f,238}$\\
$\sigma_{f,239}$\\
\hline
$\Delta\chi^{2}_{\text{NO}}$\\
$n\sigma_{\text{NO}}$\\
\end{tabular}%
\begin{tabular}[t]{c}
\\
NEOS+DANSS\\
\hline
$81.0$\\
$81$\\
$48\%$\\
\hline
$
1.29
\pm 0.03
$\\
$
0.049
\pm 0.011
$\\
$-$\\
$-$\\
$-$\\
$-$\\
$-$\\
\hline
$
0.012
\pm 0.003
$\\
$-$\\
$-$\\
$-$\\
\hline
$16.7$\\
$3.7$\\
\end{tabular}%
\begin{tabular}[t]{c}
MI$\nu_{e}$Dis\\
235+239\\
\hline
$138.5$\\
$144$\\
$61\%$\\
\hline
$
1.29
\pm 0.03
$\\
$
0.047
^{+ 0.009 }_{- 0.011 }
$\\
$
0.957
\pm 0.011
$\\
$-$\\
$
1.005
^{+ 0.034 }_{- 0.032 }
$\\
$
0.869
^{+ 0.080 }_{- 0.062 }
$\\
$
0.836
^{+ 0.075 }_{- 0.057 }
$\\
\hline
$
0.012
^{+ 0.002 }_{- 0.003 }
$\\
$
6.40
\pm 0.07
$\\
$-$\\
$
4.42
^{+ 0.15 }_{- 0.14 }
$\\
\hline
$14.9$\\
$3.4$\\
\end{tabular}%
\begin{tabular}[t]{c}
MI$\nu_{e}$Dis\\
235+238+239\\
\hline
$136.5$\\
$143$\\
$64\%$\\
\hline
$
1.29
\pm 0.03
$\\
$
0.043
^{+ 0.014 }_{- 0.009 }
$\\
$
0.970
^{+ 0.015 }_{- 0.013 }
$\\
$
0.76
^{+ 0.15 }_{- 0.16 }
$\\
$
1.056
^{+ 0.062 }_{- 0.042 }
$\\
$
0.863
^{+ 0.087 }_{- 0.061 }
$\\
$
0.854
^{+ 0.060 }_{- 0.075 }
$\\
\hline
$
0.011
^{+ 0.003 }_{- 0.002 }
$\\
$
6.49
^{+ 0.10 }_{- 0.09 }
$\\
$
7.6
^{+ 1.5 }_{- 1.6 }
$\\
$
4.65
^{+ 0.27 }_{- 0.18 }
$\\
\hline
$14.6$\\
$3.4$\\
\end{tabular}%
\caption{ \label{tab:nuedis}
Results of the fits of $\nu_{e}$ and $\bar\nu_{e}$ disappearance data:
minimum $\chi^2$ ($\chi^{2}_{\text{min}}$),
number of degrees of freedom (NDF),
goodness of fit (GoF):
best fit values of
$\Delta{m}^2_{41}$,
$\sin^22\vartheta_{ee}$,
$r_{235}$,
$r_{238}$,
$r_{239}$,
$\eta_{\text{G}}$,
$\eta_{\text{S}}$,
and those of the derived quantities
$|U_{e4}|^2$,
$\sigma_{f,235}$,
$\sigma_{f,238}$,
$\sigma_{f,239}$;
$\chi^{2}$ difference $\Delta\chi^{2}_{\text{NO}}$ between the $\chi^{2}$ of no oscillations and $\chi^{2}_{\text{min}}$, and
the resulting number of $\sigma$'s ($n\sigma_{\text{NO}}$)
for two degrees of freedom corresponding to two fitted oscillation parameters
($\sin^22\vartheta_{ee}$ and $\Delta{m}^2_{41}$).
The cross sections per fission $\sigma_{f,i}$
are expressed in units of
$10^{-43} \, \text{cm}^2 / \text{fission}$.
}
\end{table}

The statistical significance of the NEOS+DANSS
indication in favor of short-baseline $\bar\nu_{e}$ oscillations is of
$3.7\sigma$.
This value is similar to the statistical significance of the reactor and Gallium anomalies that we found in Ref.~\cite{Gariazzo:2017fdh}.
However, it is much more reliable, because it is model-independent.

Figure~\ref{fig:neos+danss-spe} shows the results of the combined fit of the
DANSS and NEOS data,
together with the $2\sigma$ allowed regions
of DANSS and NEOS.
One can see that there is a good overlap of the DANSS and NEOS allowed regions at
$\Delta{m}^2_{41} \simeq 1.3 \, \text{eV}^2$,
which determines the region preferred by the combined fit,
with the best-fit oscillation parameters in Table~\ref{tab:nuedis}.
There are also
small overlaps of the DANSS and NEOS allowed regions at
$\Delta{m}^2_{41} \simeq 0.4 \, \text{eV}^2$
and
$\Delta{m}^2_{41} \simeq 2.5 \, \text{eV}^2$
that determine two narrow islands allowed at $3\sigma$ by the combined fit.

\section{The reactor and Gallium anomalies}
\label{sec:anomalies}

As emphasized in the Section~\ref{sec:DANSS},
the DANSS and NEOS spectral ratios give indications of
short-baseline $\bar\nu_{e}$ oscillations
which are independent of the reactor flux calculation.
This indication is much more robust than those of the
reactor and Gallium anomalies,
which suffer from the dependence on the calculated reactor fluxes
and the assumed Gallium detector efficiencies.
Figure~\ref{fig:neos+danss-cmp} shows the comparison of
the NEOS+DANSS allowed regions
in the
$\sin^{2}2\vartheta_{ee}$--$\Delta{m}^{2}_{41}$ plane
with the 2 and 3$\sigma$ allowed regions of the
reactor and Gallium anomalies from Ref.~\cite{Gariazzo:2017fdh}.

From Fig.~\ref{fig:neos+danss-cmp} one can see that
the NEOS+DANSS model-independent allowed regions are compatible with the $3\sigma$
allowed regions of the reactor anomaly,
but have some tension with the $2\sigma$ allowed regions.
The tension can be quantified by the parameter goodness of fit
\cite{Maltoni:2003cu},
whose value is
$ 2\%$
($\Delta\chi^2/\text{NDF}
=
8.0
/
2$).
Hence our model-independent analysis indicate that the reactor anomaly
overestimates the $\bar\nu_{e}$ disappearance.
This is probably due to an overestimate of the reactor antineutrino fluxes.
In Section~\ref{sec:nuedis} we will obtain from the combined model-independent fit
the needed corrections to the values of the
reactor antineutrino fluxes.

Figure~\ref{fig:neos+danss-cmp} also shows that
there is a compatibility of the $3\sigma$ regions allowed by the NEOS+DANSS model-independent results
with those allowed by the Gallium anomaly,
while there is a tension of the $2\sigma$ allowed regions,
corresponding to a parameter goodness of fit of
$ 4\%$
($\Delta\chi^2/\text{NDF}
=
6.6
/
2$).
This tension suggests that the efficiencies of the
GALLEX and SAGE detectors may have been overestimated.
In the combined model-independent fit presented in Section~\ref{sec:nuedis}
we will obtain an estimate of the needed corrections to those efficiencies.

\begin{figure*}[!t]
\centering
\setlength{\tabcolsep}{0pt}
\begin{tabular}{cc}
\subfigure[]{\label{fig:rea-spe-sig}
\includegraphics*[width=0.49\linewidth]{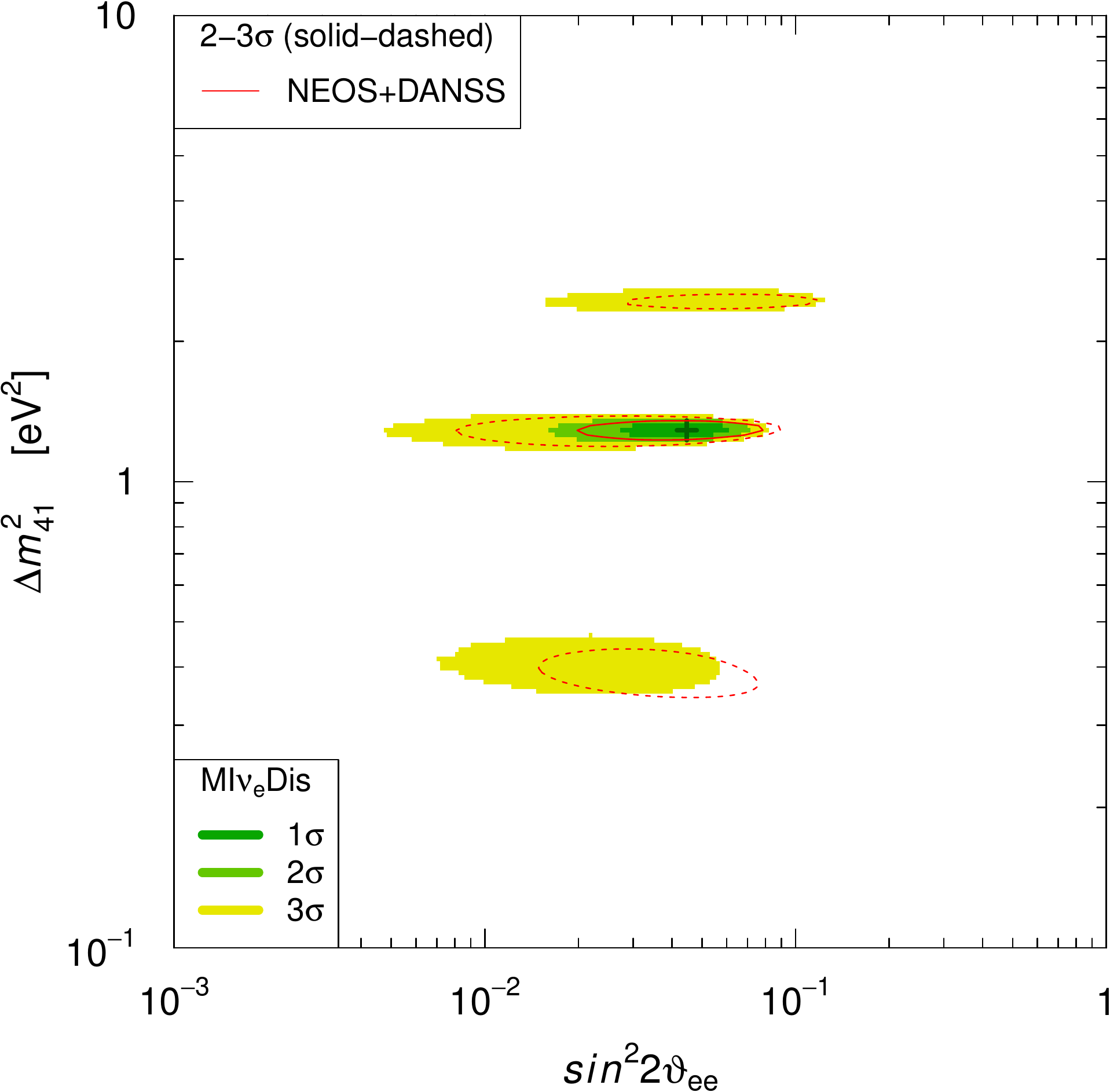}
}
&
\subfigure[]{\label{fig:fut}
\includegraphics*[width=0.49\linewidth]{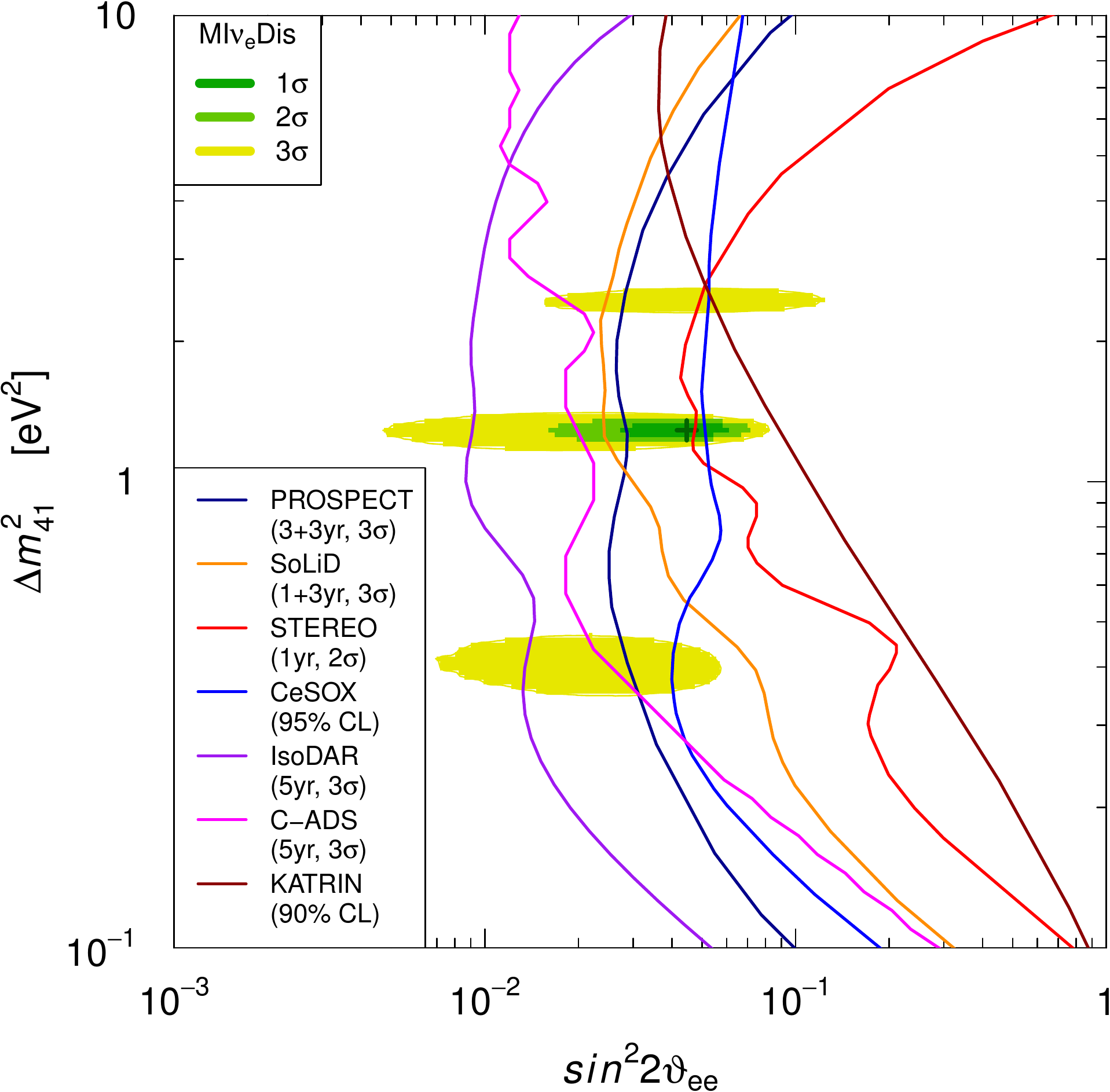}
}
\end{tabular}
\caption{ \label{fig:mid-pee}
Allowed regions in the
$\sin^{2}2\vartheta_{ee}$--$\Delta{m}^{2}_{41}$ plane obtained from the fits of
model-independent short-baseline
$\nu_{e}$ and $\bar\nu_{e}$ disappearance data (MI$\nu_{e}$Dis).
\subref{fig:rea-spe-sig}
Comparison with the regions allowed at 2 and 3$\sigma$ by the NEOS and DANSS data
(same as in Fig.~\ref{fig:neos+danss}).
\subref{fig:fut}
Comparison with the sensitivities of future experiments.
}
\end{figure*}

\begin{figure}[!t]
\centering
\includegraphics*[width=\linewidth]{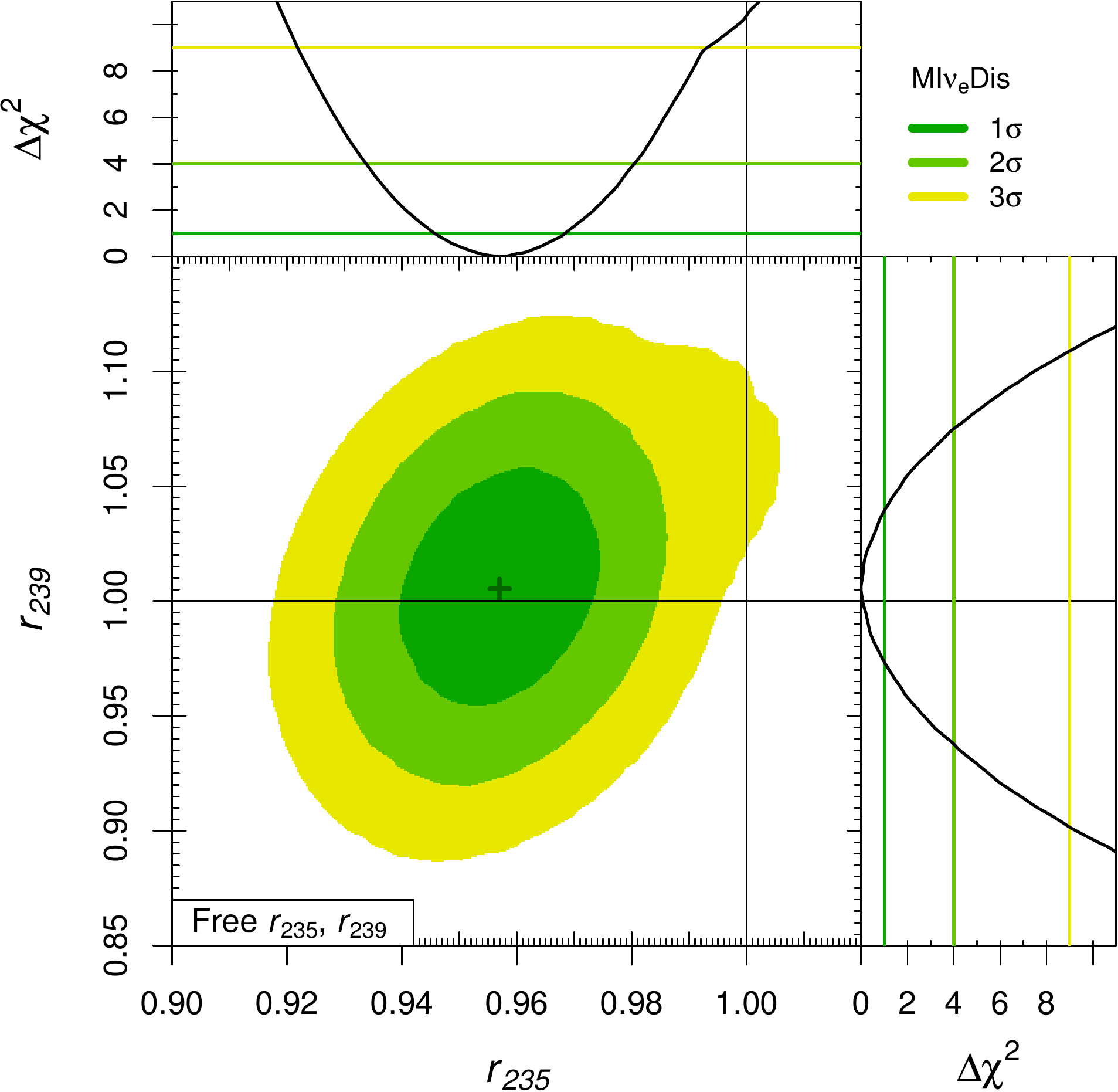}
\caption{ \label{fig:ctr-r35-r39}
Allowed regions in the
$r_{235}$--$r_{239}$ plane and marginal $\Delta\chi^{2}$'s
obtained from the fit of model-independent short-baseline
$\nu_{e}$ and $\bar\nu_{e}$ disappearance data
with free
$r_{235}$ and $r_{239}$.
The best-fit points corresponding to the $\chi^2_{\text{min}}$
in Table~\ref{tab:nuedis} are indicated by crosses.
}
\end{figure}

\section{Model-independent $\nu_{e}$ and $\bar\nu_{e}$ disappearance}
\label{sec:nuedis}

As discussed in the Section~\ref{sec:DANSS},
the combined fit of the DANSS and NEOS spectral ratios give
a model-independent indication in favor of
short-baseline $\bar\nu_{e}$ oscillations.
In particular,
it is independent of the
reactor and Gallium anomalies,
which depend, respectively,
on the reactor antineutrino flux calculation
and on the assumed efficiencies of the Gallium detectors.
In this section we present the results of a
model-independent analysis of short-baseline
$\nu_{e}$ and $\bar\nu_{e}$ disappearance data
(MI$\nu_{e}$Dis).
We considered
the DANSS and NEOS spectral ratios,
the reactor rates keeping the main reactor antineutrino fluxes as free
and the Gallium data with free efficiencies
of the GALLEX and SAGE detectors.
For completeness,
we considered also the following data which constrain neutrino oscillations in a model-independent way
and which were included in our previous analyses
\cite{Giunti:2012tn,Giunti:2012bc,Giunti:2013aea,Gariazzo:2015rra,Gariazzo:2017fdh}
of $\nu_{e}$ and $\bar\nu_{e}$ disappearance:

\begin{itemize}

\item
The ratio of the spectra measured at 40~m and 15~m from the source
in the Bugey-3 experiment \cite{Declais:1995su}.

\item
The ratio of the
KARMEN~\cite{Armbruster:1998uk}
and
LSND~\cite{Auerbach:2001hz}
$\nu_{e} + {}^{12}\text{C} \to {}^{12}\text{N}_{\text{g.s.}} + e^{-}$
scattering data at 18~m and 30~m from the source
\cite{Conrad:2011ce,Giunti:2011cp}

\end{itemize}
Let us however clarify that the contribution of these data is very small
and the results are almost independent of their inclusion in the fit.

Concerning the fit of the reactor rates,
we considered
the Daya Bay fuel evolution data~\cite{An:2017osx}
and
the rates of the following experiments
(see Table~1 of Ref.~\cite{Gariazzo:2017fdh}):
Bugey-4~\cite{Declais:1994ma},
Rovno91~\cite{Kuvshinnikov:1990ry},
Bugey-3~\cite{Declais:1995su},
Gosgen~\cite{Zacek:1986cu},
ILL~\cite{Kwon:1981ua,Hoummada:1995zz},
Krasnoyarsk87~\cite{Vidyakin:1987ue},
Krasnoyarsk94~\cite{Vidyakin:1990iz,Vidyakin:1994ut},
Rovno88~\cite{Afonin:1988gx},
SRP~\cite{Greenwood:1996pb},
Nucifer~\cite{Boireau:2015dda},
Chooz~\cite{Apollonio:2002gd},
Palo Verde~\cite{Boehm:2001ik},
RENO~\cite{RENO-AAP2016},
and
Double Chooz~\cite{DoubleChooz-private-16}.
We improved the analysis of the reactor rates presented in Ref.~\cite{Giunti:2017yid}
by taking into account the uncertainties of the reactor fission fractions.
For each experimental data point labeled with the index $a$,
we considered
the theoretical cross section per fission
(that quantifies the $\bar\nu_{e}$ detection rate)
\begin{equation}
\sigma_{f,a}^{\text{th}}
=
\sum_{i} f_{i}^{a} \overline{F}_{i}^{a} r_{i} \sigma_{f,i}^{\text{SH}}
,
\label{csfth}
\end{equation}
where $i=235,238,239,241$
and
$\sigma_{f,i}^{\text{SH}}$
are the corresponding theoretical Saclay+Huber cross sections per fission
\cite{Mueller:2011nm,Mention:2011rk,Huber:2011wv}.
The average values $\overline{F}_{i}^{a}$ of the effective fission fractions
are multiplied by the coefficients $f_{i}^{a}$
in order to take into account their uncertainties.
The coefficients $r_{i}$ allow us to consider a variation of the
$\bar\nu_{e}$ fluxes with respect to the calculated ones.
We considered the following two cases:

\begin{description}

\renewcommand{\labelenumi}{(\theenumi)}
\renewcommand{\theenumi}{\Alph{enumi}}

\item[\namedlabel{case:235+239}{235+239}]
Free $r_{235}$ and $r_{239}$ to be determined by the fit.

\item[\namedlabel{case:235+238+239}{235+238+239}]
Free $r_{235}$, $r_{238}$, and $r_{239}$ to be determined by the fit.

\end{description}
The case~\ref{case:235+239} is motivated by the fact that the
Daya Bay evolution data and the fuel composition of the other reactor experiments
constrain mainly the two major
${}^{235}\text{U}$ and ${}^{239}\text{Pu}$
fluxes
\cite{An:2017osx,Giunti:2017nww,Giunti:2017yid,Gebre:2017vmm}.
The case~\ref{case:235+238+239} is motivated by the discovery in Ref.~\cite{Gebre:2017vmm}
that also the ${}^{238}\text{U}$ flux can be loosely constrained.
On the other hand,
the ${}^{241}\text{Pu}$ flux is not constrained at all,
as we have verified through a tentative analysis with all $r_{i}$ free.

We analyzed the reactor rates with the least-squares statistic
\begin{align}
\chi^2
=
\null & \null
\sum_{a,b}
\left( P_{ee}^{a} \sum_{i} f_{i}^{a} \overline{F}_{i}^{a} r_{i} \sigma_{f,i}^{\text{SH}} - \sigma_{f,a}^{\text{exp}} \right)
(V_{\text{exp}}^{-1})_{ab}
\nonumber
\\
\null & \null
\hspace{1cm}
\times
\left( P_{ee}^{b} \sum_{j} f_{j}^{b} \overline{F}_{j}^{b} r_{j} \sigma_{f,j}^{\text{SH}} - \sigma_{f,b}^{\text{exp}} \right)
\nonumber
\\
\null & \null
+ \sum_{i,j\in\Omega}
\left( r_{i} - 1 \right)
(V_{\text{SH}}^{-1})_{ij}
\left( r_{j} - 1 \right)
\nonumber
\\
\null & \null
+ \sum_{a,b} \sum_{i,j}
\left( \dfrac{ f_{i}^{a} - 1 }{ \delta_{i}^{a} } \right)
(C_{\text{exp}}^{-1})_{ab}
(C_{f}^{-1})_{ij}
\left( \dfrac{ f_{j}^{b} - 1 }{ \delta_{j}^{b} } \right)
\nonumber
\\
\null & \null
+ \sum_{a} \lambda_{a}
\left[
1 - \sum_{i} f_{i}^{a} \overline{F}_{i}^{a}
\right]
,
\label{chi2lm}
\end{align}
where
$P_{ee}^{a}$
are the $\bar\nu_{e}$ survival probabilities,
$\sigma_{f,a}^{\text{exp}}$
are the measured cross sections per fission,
$V_{\text{exp}}$
is the experimental covariance matrix,
$V_{\text{SH}}$
is the covariance matrix of the fractional uncertainties of the Saclay-Huber theoretical calculation
of the antineutrino fluxes
from the four fissionable nuclides,
$\delta_{i}^{a}$ is the uncertainty of the fission fraction $i$ in the experiment $a$,
$C_{\text{exp}}$ is the correlation matrix of the fission fractions in the different experiments
and
$C_{f}$
is the correlation matrix of the four fission fractions.
In the second sum
$\Omega=\{238,241\}$ in the 235+239 analysis
and
$\Omega=\{241\}$ in the 235+238+239 analysis.
The coefficients $\lambda_{a}$ are Lagrange multipliers
that enforce the constraint
\begin{equation}
\sum_{i} f_{i}^{a} \overline{F}_{i}^{a}
=
\sum_{i} \overline{F}_{i}^{a}
=
1
.
\label{sff1}
\end{equation}
For experiments at commercial reactors we assumed the value
of the fission fraction uncertainty estimated by the Daya Bay collaboration~\cite{An:2016srz},
$\delta_{i}^{a}=5\%$,
neglecting possible differences on which there is no information.
On the other hand,
the fission fractions of experiments with research reactors have smaller uncertainties.
We found information on the fission fraction uncertainty
only for the Nucifer experiment~\cite{Boireau:2015dda}, where it was estimated to be 2\%.
In this experiment the enrichment in $^{235}\text{U}$ was only 19.75\%.
In the other experiments with research reactors
the uncertainty of the fission fractions should be smaller,
because the fuel is highly enriched in $^{235}\text{U}$.
Therefore, for them we assumed $\delta_{i}^{a}=1\%$.
The correlation matrix $C_{\text{exp}}$ correlates the uncertainty of the fission fraction
of experiments at the same reactor, for which
we assumed 100\% correlation.
For $C_{f}$ we used the correlation matrix
estimated by the Daya Bay collaboration in Table~2 of Ref.~\cite{An:2016srz}.

\begin{figure*}[!t]
\centering
\setlength{\tabcolsep}{0pt}
\begin{tabular}{cc}
\subfigure[]{\label{fig:mid-235+238+239-r35-r39}
\includegraphics*[width=0.49\linewidth]{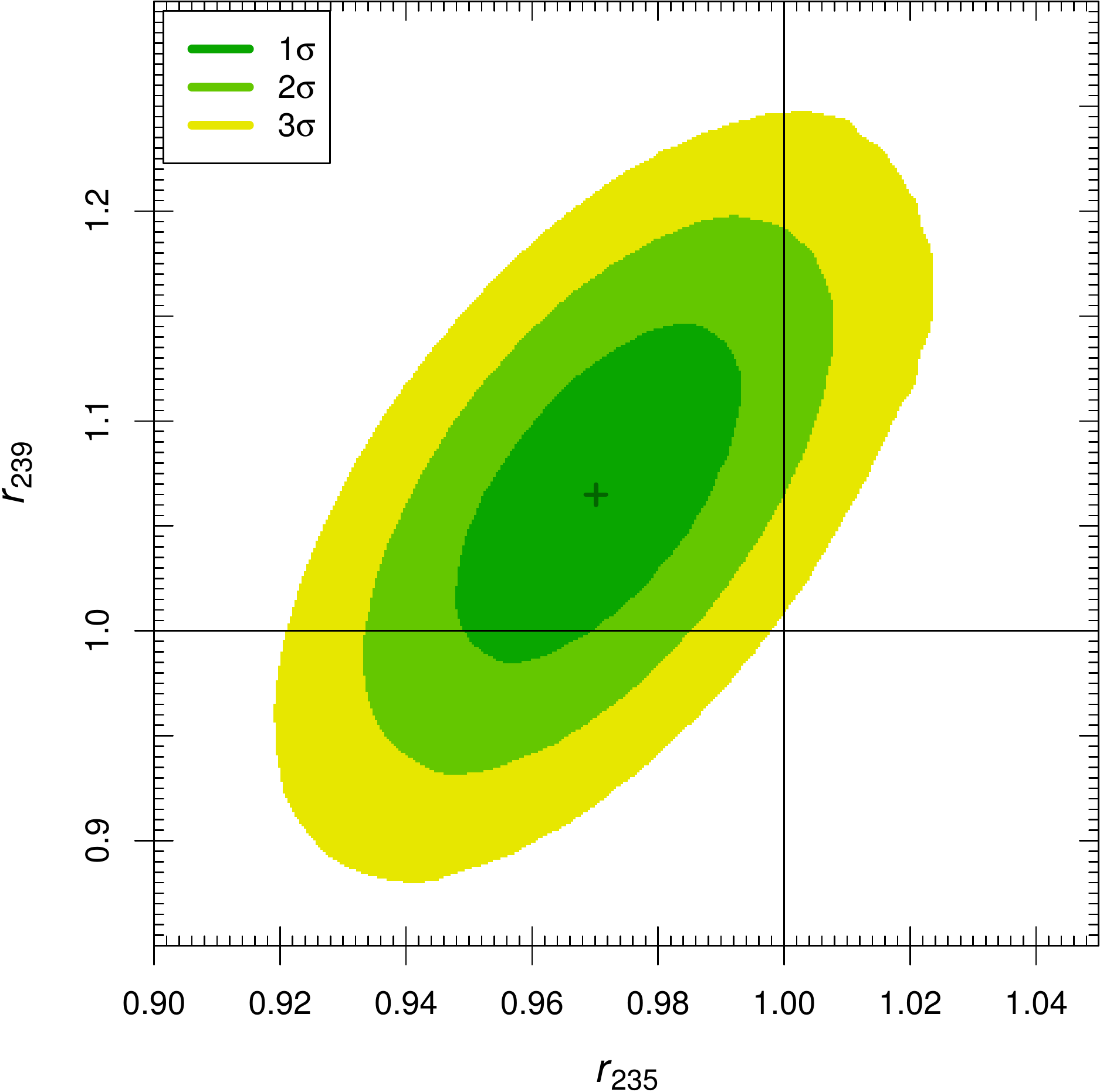}
}
&
\subfigure[]{\label{fig:mid-235+238+239-r38-r39}
\includegraphics*[width=0.49\linewidth]{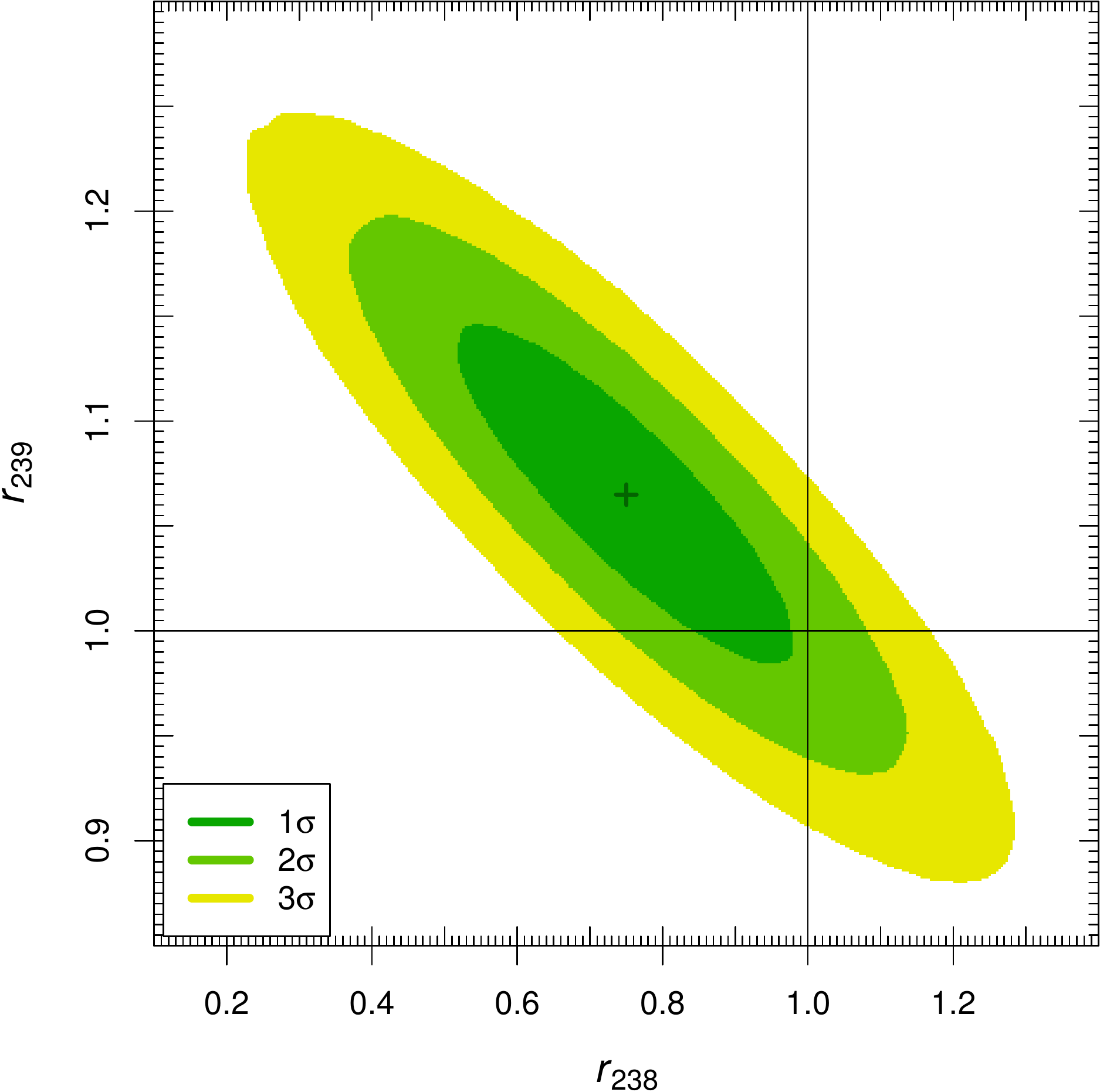}
}
\\
\subfigure[]{\label{fig:mid-235+238+239-r35-r38}
\includegraphics*[width=0.49\linewidth]{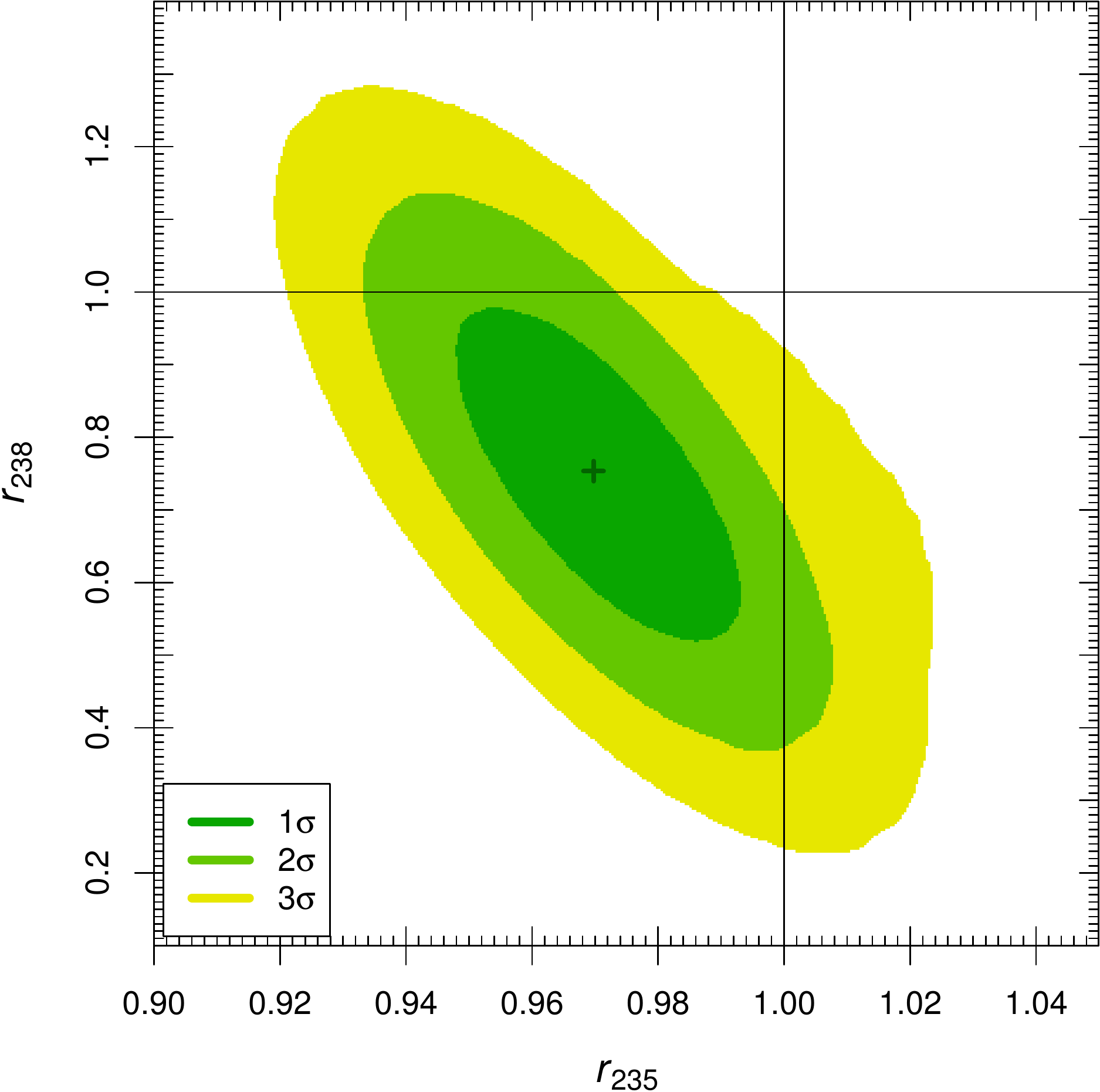}
}
&
\subfigure[]{\label{fig:mid-235+238+239-plt}
\includegraphics*[width=0.49\linewidth]{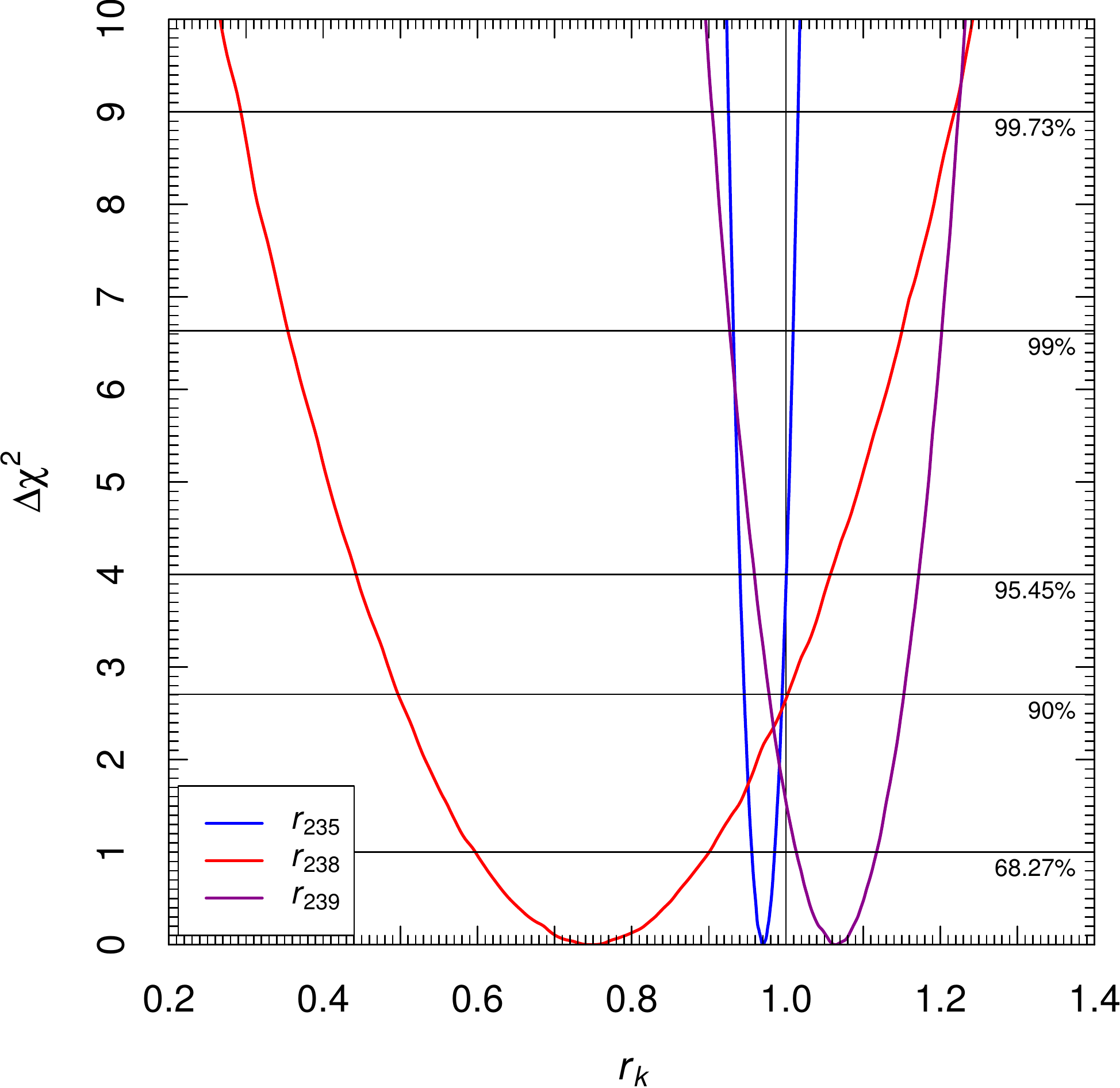}
}
\end{tabular}
\caption{ \label{fig:mid-235+238+239}
Allowed regions in the
$r_{235}$--$r_{239}$,
$r_{238}$--$r_{239}$, and
$r_{235}$--$r_{238}$
planes and marginal $\Delta\chi^{2}$'s
obtained from the fits of
all short-baseline
$\nu_{e}$ and $\bar\nu_{e}$ disappearance data
with free
$r_{235}$, $r_{238}$ and $r_{239}$.
The best-fit points corresponding to the $\chi^2_{\text{min}}$
in Table~\ref{tab:nuedis} are indicated by crosses.
}
\end{figure*}

\begin{figure}[!t]
\centering
\includegraphics*[width=\linewidth]{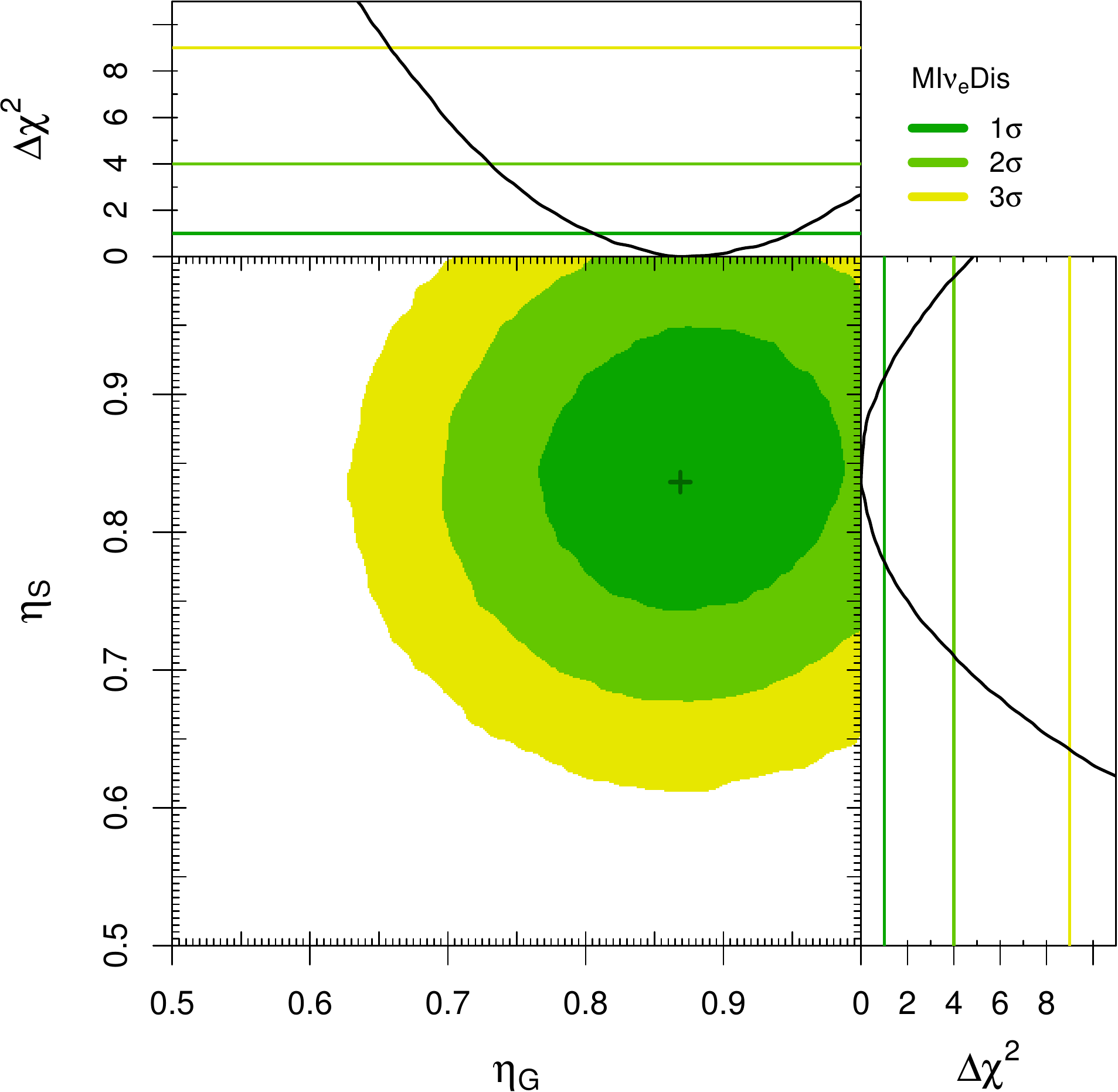}
\caption{ \label{fig:cgf}
Allowed regions in the
$\eta_{\text{G}}$--$\eta_{\text{S}}$ plane and marginal $\Delta\chi^{2}$'s
obtained from the fit of model-independent short-baseline
$\nu_{e}$ and $\bar\nu_{e}$ disappearance data.
$\eta_{\text{G}}$ and $\eta_{\text{S}}$ are, respectively,
the corrections to the efficiencies of the GALLEX and SAGE Gallium detectors.
The best-fit points corresponding to the $\chi^2_{\text{min}}$
in Table~\ref{tab:nuedis} are indicated by crosses.
}
\end{figure}

The results of the
\ref{case:235+239}
and
\ref{case:235+238+239}
fits are given in Table~\ref{tab:nuedis}.
The results for the oscillation parameters
$\sin^{2}2\vartheta_{ee}$ and $\Delta{m}^{2}_{41}$
are practically equal in the two analyses.
The allowed regions in the
$\sin^{2}2\vartheta_{ee}$--$\Delta{m}^{2}_{41}$ plane are shown in Fig.~\ref{fig:mid-pee}.
In Fig.~\ref{fig:rea-spe-sig} they are confronted with
the regions allowed at 2 and 3 $\sigma$ by the combined analysis of the NEOS and DANSS data
(shown already in Fig.~\ref{fig:neos+danss}).
One can see that the global model-independent allowed regions are mostly determined by the
NEOS and DANSS spectral data,
with small effects of the other constraints.
Indeed,
as one can see from the values in Table~\ref{tab:nuedis},
the statistical significance of short-baseline neutrino oscillations
obtained in the MI$\nu_{e}$Dis analyses
is almost the same as that obtained in the NEOS+DANSS analysis,
with a slight decrease due to the inclusion of 67 data points
that are less constraining than the NEOS and DANSS data.

Figure~\ref{fig:fut} shows the comparison of the allowed regions in the
$\sin^{2}2\vartheta_{ee}$--$\Delta{m}^{2}_{41}$ plane
with the sensitivities of the reactor experiments
PROSPECT \cite{Ashenfelter:2015uxt},
SoLid \cite{Michiels:2016qui},
STEREO \cite{Manzanillas:2017rta},
which are under way,
of the future experiments
CeSOX \cite{Borexino:2013xxa}
and
KATRIN \cite{Drexlin-NOW2016}
and of the proposed experiments
IsoDAR@KamLAND \cite{Abs:2015tbh}
and
C-ADS \cite{Ciuffoli:2015uta}.
One can see that the sensitivities of the reactor experiments
cover most of the allowed region.
Hence, they have a good chance to confirm the NEOS+DANSS indication if it is correct.
The CeSOX experiment is sensitive to the large-$\sin^{2}2\vartheta_{ee}$
parts of the allowed regions and the
KATRIN experiment is sensitive to the
large-$\sin^{2}2\vartheta_{ee}$ part of the
$3\sigma$ allowed region at
$\Delta{m}^2_{41} \simeq 0.4 \, \text{eV}^2$.
Also the proposed C-ADS experiment can cover the large-$\sin^{2}2\vartheta_{ee}$
parts of the allowed regions.
A definitive confirmation or exclusion of the NEOS+DANSS indication
may come from the
proposed IsoDAR@KamLAND experiment,
which covers almost all the $3\sigma$ allowed regions.

As explained in Section~\ref{sec:anomalies}
the model-independent analysis of the NEOS and DANSS data
indicate that the reactor anomaly
overestimates the $\bar\nu_{e}$ disappearance.
In the \ref{case:235+239} analysis the reactor anomaly is reduced through
the reduction of the ${}^{235}\text{U}$ antineutrino flux
shown by the best-fit values of $r_{235}$ in Table~\ref{tab:nuedis}.
In the \ref{case:235+238+239} analysis also
$r_{238}$
is smaller than one, but its effect is marginal
because the contribution of
$^{238}\text{U}$ to the total antineutrino flux is only about 8\%
for commercial power reactors and zero for research reactors.

Figure~\ref{fig:ctr-r35-r39}
shows the allowed regions in the
$r_{235}$--$r_{239}$ plane and the marginal $\Delta\chi^{2}$'s
obtained in the \ref{case:235+239} analysis with free
$r_{235}$ and $r_{239}$.
The correlated allowed region in the $r_{235}$--$r_{239}$ plane
has an ellipsoidal shape, except for a bulge at $3\sigma$
for large values of $r_{235}$ and $r_{239}$.
The bulge is due to the $3\sigma$ allowed region at
$\Delta{m}^2_{41} \simeq 2.5 \, \text{eV}^2$
in Fig.~\ref{fig:mid-pee},
where
$\sin^{2}2\vartheta_{ee}$
is relatively large and large values of
$r_{235}$ and $r_{239}$
are required to compensate the corresponding small
$\bar\nu_{e}$
survival probability.
From Fig.~\ref{fig:ctr-r35-r39}
and the best-fit values of
$r_{235}$--$r_{239}$
in Table~\ref{tab:nuedis},
we conclude that the indication in favor of the need for a recalculation of
$\sigma_{f,235}$ already found in Refs.~\cite{An:2017osx,Giunti:2017nww,Giunti:2017yid,Gebre:2017vmm}
is confirmed.
The value of $\sigma_{f,235}$ in Table~\ref{tab:nuedis}
is compatible, within the uncertainties,
with that found in Refs.~\cite{An:2017osx,Giunti:2017nww,Giunti:2017yid,Gebre:2017vmm}
assuming the absence of neutrino oscillations.
This is a remarkable result and we want to emphasize that
it depends on the stronger constraints on the allowed oscillation parameters due to the
DANSS and NEOS spectral ratios.
Indeed, the value of $\sigma_{f,235}$ that was found
in Ref.~\cite{Giunti:2017yid} with a fit of the reactor rates using a free $r_{235}$ and allowing neutrino oscillations is $r_{235} = 0.99 \pm 0.02$, which is compatible with our $r_{235}$ at less than 1$\sigma$.
Our analysis also confirm the results on the value of $\sigma_{f,239}$ of
Refs.~\cite{An:2017osx,Giunti:2017nww,Giunti:2017yid,Gebre:2017vmm},
which assumed the absence of neutrino oscillations, indicating that it is compatible with the theoretical prediction.

Let us now consider the \ref{case:235+238+239} analysis in which also
$r_{238}$ is free.
Figure~\ref{fig:mid-235+238+239}
shows the allowed regions in the
$r_{235}$--$r_{239}$,
$r_{238}$--$r_{239}$, and
$r_{235}$--$r_{238}$ planes and the marginal $\Delta\chi^{2}$'s.
From the figure and from Table~\ref{tab:nuedis},
one can see that the best-fit value of $r_{238}$ is rather small,
but the uncertainty of $r_{238}$ is large.
The small best-fit value of $r_{238}$ pushes the best-fit values of
$r_{235}$ and $r_{239}$ to values that are larger than those obtained in the
\ref{case:235+239} analysis.
However, we have a compatibility within the $1\sigma$ uncertainty of the values of
$\sigma_{f,235}$ and $\sigma_{f,239}$
obtained in the
\ref{case:235+239} and \ref{case:235+238+239} analyses.
In particular,
as one can see from the marginal $\Delta\chi^{2}$ in Fig.~\ref{fig:mid-235+238+239-plt},
in the \ref{case:235+238+239} analysis
the theoretical value of $\sigma_{f,235}$ is disfavored at about $2\sigma$.
This result is less strong than the one found in the 235+239 case, where the theoretical value is disfavored at more than 3$\sigma$
(see the marginal $\Delta\chi^{2}$ in Fig.~\ref{fig:ctr-r35-r39}),
but it is still a significant finding.

For the ${}^{238}\text{U}$ antineutrino flux
we find a relatively strong suppression
($
r_{238}
=
0.76
^{+ 0.15 }_{- 0.16 }
$)
but the uncertainty is large and the theoretical flux calculated in Ref.~\cite{Mueller:2011nm}
is allowed at less than 2$\sigma$.
A similar suppression of the ${}^{238}\text{U}$ antineutrino flux
with a large uncertainty
was found in Ref.~\cite{Gebre:2017vmm}
assuming the absence of neutrino oscillations.
Let us remind that the ${}^{238}\text{U}$ antineutrino flux
is the only one that was calculated in Ref.~\cite{Mueller:2011nm}
``ab initio'' using the nuclear databases.
The corresponding $\beta$ spectrum was measured afterwards in Ref.~\cite{Haag:2013raa}.
The resulting converted $\nu_{e}$ spectrum is larger than the calculated spectrum for
$E_{\nu} \lesssim 3.5 \, \text{MeV}$
and smaller for
$4 \, \text{MeV} \lesssim E_{\nu} \lesssim 6.5 \, \text{MeV}$,
albeit with large uncertainties
(see Fig.~2 of Ref.~\cite{Haag:2013raa}).

In our analysis we multiply the predicted rate of the GALLEX and SAGE experiment
by the free coefficients
$\eta_{\text{G}}$ and $\eta_{\text{S}}$,
respectively,
which are the corrections to the detector efficiencies
needed to fit the Gallium data in the MI$\nu_{e}$Dis fit.
The best-fit values and uncertainties of
$\eta_{\text{G}}$ and $\eta_{\text{S}}$
are given in Table~\ref{tab:nuedis}.
Figure~\ref{fig:cgf} shows
the allowed regions in the
$\eta_{\text{G}}$--$\eta_{\text{S}}$ plane and the marginal $\Delta\chi^{2}$'s.
One can see that
$\eta_{\text{G}}$ and $\eta_{\text{S}}$
are practically uncorrelated and
the fit indicates that they are smaller than one, but the uncertainties are large.
Hence, we cannot make a definite conclusion about the
efficiencies of the GALLEX and SAGE detectors,
but we think that it would be appropriate to take into account their uncertainties
in the analysis of the GALLEX and SAGE
solar neutrino data.

\section{Conclusions}
\label{sec:conclusions}

In this paper we have considered the new model-independent indications
in favor of short-baseline $\bar\nu_{e}$ oscillations
found in the DANSS experiment \cite{DANSS-171201}
and we have shown that they reinforce the model-independent indication
in favor of short-baseline $\bar\nu_{e}$ oscillations
found in the late 2016 in the NEOS experiment \cite{Ko:2016owz}.
In the framework of 3+1 active-sterile neutrino mixing,
the combined analysis of the DANSS and NEOS spectral ratios
constrain at $2\sigma$ the two mixing parameters
$\sin^{2}2\vartheta_{ee}$ and $\Delta{m}^{2}_{41}$
to a narrow-$\Delta{m}^{2}_{41}$ island at
$\Delta{m}^2_{41} \simeq 1.3 \, \text{eV}^2$,
with
$
\sin^{2}2\vartheta_{ee}
=
0.049
\pm 0.023
$ ($2\sigma$).
If we consider the $3\sigma$ allowed regions,
there are also two islands at
$\Delta{m}^2_{41} \simeq 0.4 \, \text{eV}^2$
and
$\Delta{m}^2_{41} \simeq 2.5 \, \text{eV}^2$.
The statistical significance of the model-independent NEOS+DANSS
indication in favor of short-baseline $\bar\nu_{e}$ oscillations is of
$3.7\sigma$.

We have shown that the DANSS and NEOS indication of short-baseline $\bar\nu_{e}$ oscillations
is in tension with the reactor anomaly and with the Gallium anomaly.
However,
since the oscillation parameters are determined in a model-independent way by the NEOS and DANSS data,
it is possible to analyze the data on the reactor and Gallium anomaly
in a model-independent way,
considering as free the main reactor antineutrino fluxes and
the efficiencies of the GALLEX and SAGE detectors.
We presented the results of two analyses of this type:
one with free $^{235}\text{U}$ and $^{239}\text{Pu}$ fluxes
and one in which also the $^{238}\text{U}$ is free.
In these global model-independent analyses of short-baseline
$\nu_{e}$ and $\bar\nu_{e}$ disappearance data we took into account also
other data which constrain neutrino oscillations in a model-independent way:
the Bugey-3 spectral ratio \cite{Declais:1995su}
and the ratio of the
KARMEN~\cite{Armbruster:1998uk}
and
LSND~\cite{Auerbach:2001hz}
$\nu_{e} + {}^{12}\text{C} \to {}^{12}\text{N}_{\text{g.s.}} + e^{-}$
scattering data.
We found that the strong constraints on short-baseline neutrino oscillations
obtained from the DANSS and NEOS spectral ratios
persist in the global model-independent analyses
and allow us to obtain simultaneous information on
neutrino oscillations, on the reactor antineutrino fluxes and on the efficiencies of the Gallium detectors.
In particular,
we confirm the indication in favor of the need for a recalculation of
the $^{235}\text{U}$
cross sections per fission
found in Refs.~\cite{An:2017osx,Giunti:2017nww,Giunti:2017yid,Gebre:2017vmm}
assuming the absence of neutrino oscillations.

\begin{acknowledgments}
The work of S.G.\ is supported by the Spanish grants FPA2017-85216-P, SEV-2014-0398 (MINECO), PROMETEOII/2014/084 (Generalitat Valenciana), and has received funding from the European Union's Horizon 2020
research and innovation programme under the Marie Sk\l{}odowska-Curie grant agreement No 796941.
The work of Y.F.L.\ is supported
in part by the National Natural Science Foundation of China under Grant No. 11305193, by the Strategic Priority Research Program of the Chinese Academy of Sciences under Grant No. XDA10010100, and the CAS Center for Excellence in Particle Physics (CCEPP).
\end{acknowledgments}

%

\end{document}